# Observations of Comets C/2007 D1 (LINEAR), C/2007 D3 (LINEAR), C/2010 G3 (WISE), C/2010 S1 (LINEAR), and C/2012 K6 (McNaught) at large heliocentric distances

Oleksandra Ivanova[1,*], Luboš Neslušan[2], Zuzana Seman Krišandová[2], Ján Svoreň[2], Pavlo Korsun[1], Viktor Afanasiev[3], Volodymyr Reshetnyk[1,4], Maxim Andreev[5,6]

[1-] *Main Astronomical Observatory of NAS of Ukraine, Akademika Zabolotnoho 27, 03680 Kyiv, Ukraine*

[*] *Corresponding Author. E-mail address: sandra@mao.kiev.ua*

[2] *Astronomical Institute of the Slovak Academy of Sciences, SK-05960 Tatranská Lomnica, Slovak Republic*

[3] *Special Astrophysical Observatory, RAS, 369167 Nizhnij Arkhyz, Russia*

[4] *Taras Shevchenko National University of Kyiv, 03022 Kyiv, Glushkova ave. 4, Ukraine*

[5] *International Center for Astronomical, Medical and Ecological Research, NAS of Ukraine, Akademika Zabolotnoho 27, 03680 Kyiv, Ukraine*

[6] *Terskol Branch of the Institute of Astronomy of RAS - 81 Elbrus ave., ap. 33, Tyrnyauz Kabardino-Balkaria Republic, 361623 Russian Federation*

**ABSTRACT**

Photometric and spectroscopic observations of five nearly parabolic comets with eccentricity larger than 0.99 at heliocentric distances greater than 4 AU were performed. No molecular emission was observed for any studied comet and the entire cometary activity in all cases was attributed to dust production. Upper limits of the gas production rates for the main neutral molecules in the cometary comae were calculated. The derived values of dust apparent magnitudes were used to estimate the upper limit of the geometric cross-section of cometary nuclei (upper limits of radii range from 2 km to 28 km). Due to the poor sublimation of water ice at these distances from the Sun, other mechanisms triggering activity in comets are discussed.

## 1. Introduction

As already pointed out by Roemer (1962), bright comets are the basic source of our knowledge about the physics of these celestial objects. The bright comets are mostly small objects observed at relatively short heliocentric distances during their favourable configuration relative to the Sun and the Earth. This fact means that the typical properties derived for them are not necessarily identical with the properties of larger objects observed at larger distances.

Progress in the development of the technologies of modern light detectors and a participation of a number of large telescopes has led to a great increase in the number of observations of distant comets with perihelion distances larger than 4 AU. Unfortunately, the studies of the comets, which are active beyond the orbit of Jupiter, are episodic. Only a small number of comets and centaurs active at large heliocentric distances have been studied (Korsun and Chörny 2003; Bauer et al., 2003; Tozzi et al. 2003; Jewitt 2009; Lowry and Fitzsimmons, 2005; Meech et al. 2009; Korsun et al. 2008, 2010; Mazzotta Epifani et al. 2010, 2014; Shi et al. 2014; Rousselot et al., 2014). Differences in the levels of activity, in the relative abundances of dust and different coma morphologies have been observed. In addition, the differences between the individual objects were large even for similar geometrical conditions.

Analysis of observations of distant comets allows us to study various physical mechanisms triggering the activity at large heliocentric distances. To explain the cometary activity at large heliocentric distances a few mechanisms (see, e.g., review by Meech and Svoren, 2004; Gronkowski 2005) have been proposed. The most popular sources of the energy required to explain the activity are: the sublimation of more volatile compounds like CO or $CO_2$ (Houpis and Mendis, 1981; Prialnik and Bar-Nun, 1992; Hughes, 1992), the transition phase between amorphous and crystalline water ice (Prialnik, 1992; Gronkowski and Smela, 1998; De Sanctis et al., 2002), polymerisation of HCN (Rettig et al., 1992), and the annealing of amorphous water ice (Meech et al. 2009).

Observations of distant comets by different methods (e.g., photometry, spectroscopy, and polarimetry) can be used to determine the sizes of cometary nuclei (Svoren, 1983), to study the brightness evolution and dust composition of cometary comae (Meech et al., 2009), as well as to detect gas emissions above the reflected solar continuum (Larson, 1980; Cochran et al., 1980, 1982; Bockelée-Morvan et al., 2001; Rauer et al., 2003; Cook et al., 2005; Korsun et al., 2006, 2008).

To enlarge the set of the comets studied at relatively large heliocentric distances, in Sects. 2 and 3 we describe the observations and their analysis five comets with perihelia near or, mostly, beyond the snow line. The determined physical characteristics are given in related tables and the relations between some parameters are shown in several figures. Observations of the comets were performed utilizing the 6-meter telescope of the Special Astrophysical Observatory and the 0.6-meter telescope of the Peak Terskol Observatory, both located in Russia.

## 2. Observations and reduction

The data set of distant active comets presented in this paper was obtained within the period from 2008 to 2014. The images and spectra were obtained during several observing runs at two different sites: the Special Astrophysical Observatory (SAO RAS, Russia) and the Peak Terskol Observatory. We obtained images and spectra for a number of comets within the framework of a program of optical spectroscopic and photometric investigations of distant active comets. Table 1 summarizes the data and the geometric circumstances at the times of the observations. A general description of the comets and orbit planes is presented in Fig.1.

**Table 1.** Log of observations.

| Object | Date, UT | Exp. Times (s) | Orb.[a] | $r$[b], AU | $\Delta$[c], AU | Phase angle, (º) | Filter/ Spectrum | Telescope |
|---|---|---|---|---|---|---|---|---|
| C/2007 D1 (LINEAR) | 2008-03-14 | 7×60 5×60 4×900 | O | 8.93 | 7.94 | 1.0 | V R spectra | 6-m, SCORPIO |
| C/2007 D3 (LINEAR) | 2008-12-04 | 7×60 4×900 | O | 6.62 | 6.51 | 8.6 | V spectra | 6-m, SCORPIO |
| C/2010 G3 (WISE) | 2011-03-29 | 6×90 6×60 4×900 | O | 5.60 | 5.19 | 9.7 | V R spectra | 6-m SCORPIO-2 |
| C/2010 S1 (LINEAR) | 2011-11-25 | 10×30 7×30 5×30 4×900 | I | 7.00 | 6.52 | 7.3 | B V R spectra | 6-m, SCORPIO-2 |
| C/2012 K6 (McNaught) | 2014-02-13 | 10×180 11×120 11×120 | O | 4.16 | 3.52 | 11.3 | B V R | 0.6-m Zeiss |
|  | 2014-02-23 | 7×180 5×180 | O | 4.22 | 3.45 | 9.4 | B V | 0.6-m Zeiss |

[a] Orbital arc: I is inbound leg of orbit (pre-perihelion); O is outbound leg of orbit (post-perihelion)
[b] $r$ is heliocentric distance
[c] $\Delta$ is geocentric distance.

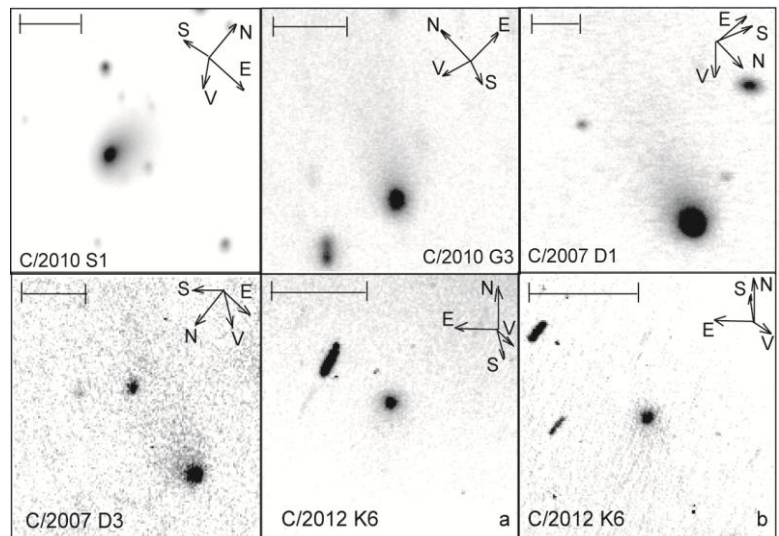

**Fig.1.** The positions of a given comet for the observational dates. The summarized images of the comets obtained with the filter $R$, celestial north (N), east (E), the motion (V) and sunward directions (S) are also shown. The scale bar represents $5·10^5$ km perpendicular to the line of sight.
3

## 2.1 *Observations at 6 m telescope of SAO RAS*

The observations of comets C/2007 D1 (LINEAR), C/2007 D3 (LINEAR), C/2010 G3 (WISE) and C/2010 S1 (LINEAR) were made with the 6 meter telescope of the Special Astrophysical Observatory (SAO RAS). The heliocentric distances of the comets in the observation period ranged from 4.22 AU to 8.93 AU. More detailed information about the observations of the comets is given in Table 1. The observations were made using the universal SCORPIO and SCORPIO-2 focal reducers mounted in the primary focus of the 6-m BTA telescope of the SAO RAS (Afanasiev and Moiseev, 2011). We used the photometric and spectroscopic modes of the focal reducers. An EEV 42-40 chip of 2048×2048 pixels was employed as the detector with the reducer SCORPIO. The SCORPIO-2 device was used with a CCD radiation detector E2V 42–90. The size of an image was 2048×2048 pixels. The image scale was 0.18 arcsec/pix and the full field of view of the detectors was 6.1×6.1 arcmin.

We applied 2×2 and 2×1 binning to the photometric and spectroscopic frames respectively during the observations. The reduction of the raw data including bias subtraction and flat-field corrections was made. The telescope was tracked on the comet to compensate its apparent motion during the exposure. The frames with morning sky were obtained to create the averaged flat-field image for the photometric data, while a smoothed spectrum of an incandescent lamp was observed for the spectral data. We used routine sky (http://www.astro.washington.edu/docs/idl/cgi-bin/getpro/library01.html?SKY) of the IDL library (Goddard Space Flight Center) to calculate the sky background count (Landsman, 1993). Observed frames were cleaned from cosmic events. The recorded events were removed automatically when we computed a composite image from the individual ones. For this purpose, the robust routine of IDL library (http://www.astro.washington.edu/docs/idl/cgi-bin/getpro/library30.html?ROBOMEAN) was applied to the stacked images.

The spectroscopic data were obtained using the VPHG1200B grism (in 2008) and VPHG940@600 grism (in 2011) in combination with a long, narrow-slit mask having dimensions of 1.0″ × 6.1′. The spectral resolution of the spectra was defined by the width of the slit and was about 5 Å. The photometric data of the comets were obtained in Johnson-Cousins broadband filters *BVR*. We obtained the observations in 2008 in a crowded field with seeing being stable to about 1.8″, while the observing conditions were much better in 2011 (seeing ~1.3″). All nights were photometric.

We observed the spectrophotometric standard stars BD+28d4211, HZ44, and GD108 (Oke, 1990) to make the absolute calibration of the observed spectra and photometry of the comets at 6-m telescope SAO RAS. To perform an absolute flux calibration of the comet images obtained at 0.6 m telescope of Peak Terskol Observatory, the field stars were used. The background stars were identified on a comet image. We selected all bright stars as photometric standards with the signal-to-noise ratio greater than 100. The stellar magnitudes of the standard stars were taken from the catalog NOMAD (Zacharias, et al., 2005) (http://vizier.u-strasbg.fr/viz-bin/VizieR). We used the correspondence between instrumental and catalogue magnitudes in each filter to derive the transformation coefficient from the instrumental to the B, V, and R system. After studying the star in the field of view, we used for our study only those standard stars, which were not variable. For each observed comet, we obtained the integrated magnitude as a function of the aperture.

### 2.2 *Observations at 0.6 m telescope of Peak Terskol Observatory*

The comet C/2012 K6 (McNaught) was observed with the 60-cm Zeiss telescope of the Peak Terskol Observatory (IC AMER) on February 13 and 23, 2014. We observed the comet when it was at heliocentric distances 4.16 AU and 4.22 AU after its perihelion passage. A brief summary of our observations of the comet is presented in Table 1.

The CCD SBIG STL 1001 camera was used as a detector. The size of the image was 1024 × 1024 pixels. The full field view of the CCD was 10.6×10.6 arcmin, and the image scale was 1.24 arcsec per pixel. The seeing was ~1.9″.

The photometric images of the comet were obtained in Johnson *B*, *V,* and *R* broad-band filters centred at 4330 Å, 5450 Å, and 6460 Å, respectively. A binning of 2×2 was applied during the observations. After applying the standard procedures of bias subtraction and flat-fielding, all individual frames were stacked together and summed to increase the signal-to-noise ratio.

## 3. Analysis of the results

### 3.1 Spectroscopy

To reduce the long-slit spectra observed with SCORPIO we used the package Scorpio2k.lib developed at SAO RAN. The package enables making the basic photometric reductions, such as bias removing, flat-fielding, geometry correction, and linearization of the scale along dispersion. The night sky spectrum existing in the frames with the cometary spectra was removed by measuring its level in each column over the zones that were free of the cometary coma. The photometric calibration of the cometary spectra was made using observed spectra of the spectrophotometric standard stars and spectral behavior of atmospheric extinction that was taken from Kartasheva and Chunakova (1978). A median frame for each considered comet was composed from a number of individual images with spectra. The last procedure also eliminates cosmic rays. Since the observed spectra were obtained from the faint objects we collapsed them in spatial direction to increase the signal-to-noise ratio of the spectra to be analysed. To search for possible molecular emissions we need to remove continuum from the observed spectra. In fitting the continuum we used a solar spectrum, which was taken from Neckel and Labs (1984). Convolving the solar spectrum with the appropriate instrumental profile we degraded it to the resolution of our observations. The proper spectral behavior was gained multiplying the solar spectrum by a polynomial. The observed spectra and fitted continua for the comets observed at SAO RAN are displayed in Fig. 2.

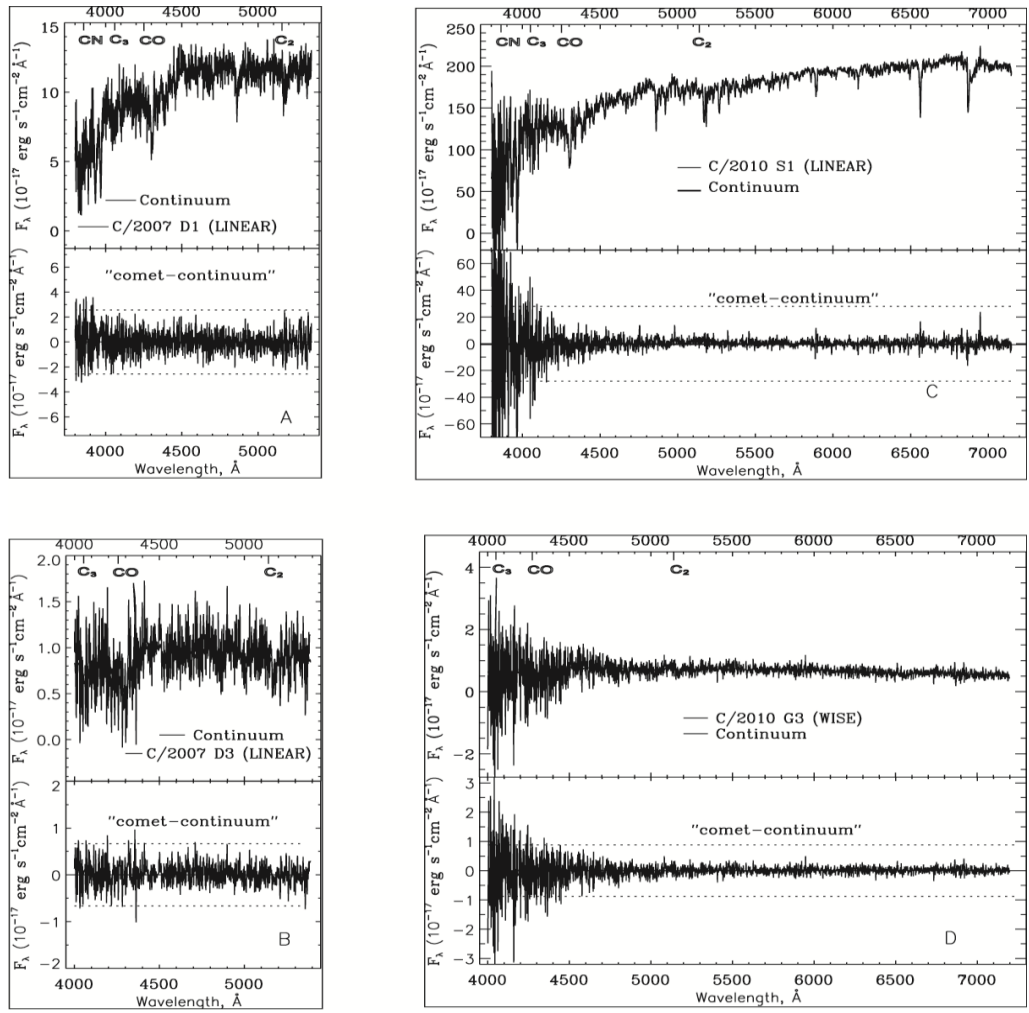

**Fig. 2.** Spectra of four distant comets. The spectra of C/2007 D1 (LINEAR), C/2007 D3 (LINEAR), C/2010 G3 (WISE), and C/2010 S1 (LINEAR) are shown in four frames. The observed spectra and superimposed fitted continua are depicted in the upper panels of each frame. The results of subtraction of the calculated continua from the observed spectra are depicted in the lower panels of each frame. The wavelength limits of the spectra are defined by the grisms used and by the level of the detected signal. The dotted lines correspond to the ±3σ levels of the residual data. The positions of the most intensive spectral lines of CN, $C_3$, $CO^+$, and $C_2$, which are typically seen in cometary, but not occurring in our spectra, are indicated at the upper border of each plot.

Subtracting the calculated continuum from the observed spectrum, we obtained a residual signal, in which the potential emission features would be detected. The result of the subtraction is also displayed in Fig. 2 (lower panels in the ABCD frames). The 3σ level was used as a criterion to search for the emissions in the noisy residuals. The careful inspection of the data shows that no regular feature exceeds the 3σ level. The data displayed in Fig.2 (lower panels of frames C and D) with the values above the specified level do not show any regular features. We consider it as noise that is determined at the low spectral sensitivity of the E2V 42–90 detector used in the blue wavelength region.

The polynomials used to fit the observed continua can also be used to measure the reddening effect when the slit was oriented along the parallactic angle. In our cases the slit direction deviated by 45°-103° relative to the mean parallactic angle and we do not make any conclusions on the spectral reddening gradient.

We concluded that we have detected only the continuum and not detected any emission in the spectra of four distant comets. A few spectra for distant comets (with $q > 5$ AU) were published so far. The spectra of well-known periodic comet 29P/Schwassmann–Wachmann 1 show the CN and ionic emissions at distances of ~6 AU from the Sun (Cochran et al., 1980; Cochran and Cochran, 1991). Ionic emissions were also detected in the spectra of the long period comet C/2002 VQ94 (LINEAR) at the distance of 8.36 AU from the Sun (Korsun et al., 2014). Nevertheless, no emission exceeding 3σ level was found in the spectra of comets C/2003 WT42 (LINEAR) (Korsun et al., 2010) and C/2006 S3 (LONEOS) (Rousselot et al., 2014).

We calculated upper limits of the emission fluxes of $CO^+$ and upper limits of the production rates of the neutrals, which could be detected at large heliocentric distances. In order to calculate the relative fluxes we used an approach similar to that presented by Rousselot et al. (2014). The upper limits of the production rates of the neutrals were derived using the Haser model (Haser, 1957). The model parameters for the neutrals were taken from Langland-Shula and Smith (2011). G-factor for CN, which depends on the heliocentric velocity of the comet, was taken from Schleicher (2010). The resulting upper limits are presented in Table 2.

**Table 2.** Upper limits for the main molecules in the cometary comae.

| Molecule | Central wavelength and spectral range, Å/Δλ | Amplitude=σ, $10^{-17}$ erg s$^{-1}$ cm$^{-2}$ Å$^{-1}$ | Flux, $10^{-16}$ erg s$^{-1}$ cm$^{-2}$ | Aperture, $l \times h$ arcsec$^2$ | Gas production rate**, $10^{23}$ mol s$^{-1}$ |
|---|---|---|---|---|---|
| **C/2007 D1 (LINEAR)** | | | | | |
| CN | 3870/62 | 1.43 | <0.77 | 1×22 | <15.62 |
| $C_3$ | 4062/62 | 1.02 | <0.55 | | <0.32 |
| $CO^+$ | 4266/64 | 0.63 | <0.34 | | |
| $C_2$ | 5141/118 | 0.81 | <0.44 | | <5.87 |
| **C/2007 D3 (LINEAR)** | | | | | |
| $C_3$ | 4062/62 | 0.33 | <0.18 | 1×18 | <0.06 |
| $CO^+$ | 4266/64 | 0.24 | <0.13 | | |
| $C_2$ | 5141/118 | 0.16 | <0.09 | | <0.64 |
| **C/2010 G3 (WISE)** | | | | | |
| $C_3$ | 4062/62 | 1.50 | <0.85 | 1×24 | <0.15 |
| $CO^+$ | 4266/64 | 0.74 | <0.42 | | |
| $C_2$ | 5141/118 | 0.15 | <0.08 | | <0.27 |
| **C/2010 S1 (LINEAR)** | | | | | |
| CN | 3870/62 | 57.77 | <31.51 | 1×37 | <234.17* |
| $C_3$ | 4062/62 | 22.51 | <12.72 | | <2.77 |
| $CO^+$ | 4266/64 | 7.34 | <4.15 | | |
| $C_2$ | 5141/118 | 3.03 | <1.71 | | <7.36 |

\* The high value of the gas production rate of CN is caused by the high level of noise within the examined wavelength region.

\*\* The gas production rates were calculated for an effective annular aperture with diameter of $D = \sqrt{4hl/\pi}$, where $h$ and $l$ are height and width of the slit respectively.

### 3.2 Photometry

All comets were observed at heliocentric distances larger than 4 AU. The spectral observations of a majority of distant comets did not reveal molecular emissions seen above the reflected solar continuum. We used broadband filters for the analysis of the cometary dust environment. We obtained the magnitude using relation

$$m_c = -2.5 \cdot \log\left[\frac{I_c(\lambda)}{I_s(\lambda)}\right] + m_{st} - 2.5 \cdot \log(P(\lambda)) \cdot \Delta M , \qquad (1)$$

where $m_{st}$ is the magnitude of the standard star; $I_s(\lambda)$ and $I_c(\lambda)$ are the measured fluxes of the star and the comet in counts, respectively; $P(\lambda)$ is the sky transparency that depends on the wavelength, and $\Delta M$ is the difference between the cometary and stellar air masses. Because the field stars were used for calibration, the sky transparency is not taken into account.

Absolute magnitudes of the comets were corrected to $r = \Delta = 1$ AU and phase angle $\alpha = 0º$ using equation

$$m_R(1,1,0) = m_R - 5 \cdot \log(r \cdot \Delta) - \beta \cdot \alpha, \qquad (2)$$

where $r$ and $\Delta$ are, respectively, the heliocentric and geocentric distances of the comet (in AU); $m_R$ is R magnitude of the comet; $\alpha$ is the phase angle of the comet (in degree); $\beta = 0.04$ is the linear phase coefficient in magnitudes per degree (Snodgrass et al., 2008). The results for our comets are given in Table 3.

**Table 3.** Photometry, reduced magnitudes, and colors of the comets.

| Comet | DATE, UT | r, AU | $m_R(1,1,0)$ | ρ, arcsec | $m_c^a$ | Color index BV* | VR* |
|---|---|---|---|---|---|---|---|
| C/2007 D1(LINEAR) | 2008/03/14 | 8.93 | 12.95±0.03 | 1.8 | 19.36±0.08 | - | 0.59±0.09 |
|  |  |  | 12.33±0.03 | 2.2 | 19.10±0.08 |  | 0.57±0.09 |
|  |  |  | 11.04±0.03 | 3.3 | 18.57±0.08 |  | 0.46±0.09 |
| C/2007 D3 (LINEAR) | 2008/12/04 | 6.62 | - | 1.8 | 20.89±0.07 | - | - |
|  |  |  |  | 2.2 | 20.68±0.06 |  |  |
|  |  |  |  | 3.3 | 20.29±0.05 |  |  |
| C/2010 G3 (WISE) | 2011/03/29 | 5.6 | 14.74±0.01 | 1.8 | 20.36±0.07 | - | 0.61±0.07 |
|  |  |  | 14.08±0.01 | 2.2 | 20.09±0.07 |  | 0.61±0.07 |
|  |  |  | 12.73±0.01 | 3.3 | 19.59±0.06 |  | 0.56±0.06 |
| C/2010 S1 (LINEAR) | 2011/11/25 | 7.0 | 11.89±0.01 | 1.8 | 16.90±0.08 | 0.62±0.11 | 0.49±0.08 |
|  |  |  | 10.01±0.01 | 2.2 | 16.59±0.07 | 0.58±0.10 | 0.48±0.07 |
|  |  |  | 8.64±0.01 | 3.3 | 15.99±0.05 | 0.64±0.09 | 0.46±0.06 |
| C/2012 K6 (MgNaught) | 2014/02/13 | 4.16 | 12.59±0.04 | 6.2 | 17.63±0.08 | 0.63±0.14 | 0.60±0.09 |
|  |  |  | 12.02±0.03 | 7.5 | 17.44±0.07 | 0.62±0.13 | 0.59±0.08 |
|  | 2014/02/23 | 4.22 | - | 6.2 | 17.79±0.08 | 0.78±0.14 | - |
|  |  |  |  | 7.5 | 17.65±0.07 | 0.75±0.13 |  |

a- magnitude of the comets obtained through V filter

*- estimated accuracy of determining the brightness of standard stars for catalogue NOMAD is about a few hundredths of magnitude.

A majority of our comets exhibit low activity. To calculate their dust production rate, we therefore used the method proposed by Jewitt (2009), which involves an estimate of the apparent magnitude of the cometary dust coma in an annular aperture, in particular

$$m_d = -2.5 \cdot \log(10^{-0.4 m_2} - 10^{-0.4 m_1}), \qquad (3)$$

where $m_1$ and $m_2$ are the magnitudes corresponding to the aperture radii $\rho_1$ and $\rho_2$; the values of the latter were taken to be $\rho_1$ =1.8 arcsec and $\rho_2$=2.4 arcsec.

The apparent magnitude can be used to estimate the upper limit of the geometric cross-section of a cometary nucleus (Russell, 1916; Jewitt, 1991). For this purpose, we used the relation

$$p(\lambda) \cdot \Phi(\alpha) \cdot C_d = 2.25 \cdot 10^{22} \cdot \pi \cdot r^2 \cdot \Delta^2 \cdot 10^{-0.4(m_d - m_{SUN})}, \qquad (4)$$

where $p(\lambda)$ is the geometric grain albedo for the given wavelength referring to the mix of grain size; $\Phi(\alpha) = 10^{-0.4 \cdot \alpha \cdot \beta}$ is the phase function dependence on phase angle $\alpha$; $C_d$ is the geometrical cross-section of the nucleus in $m^2$; $m_{SUN}$ is the magnitude of the Sun (the latter being -26.09, -27.74, and -27.26 for the B, V, and R bands, respectively).

In our calculation we used the three values of the geometric albedo $p(\lambda)$. The first adopted value $p(\lambda)$ = 0.1 is related to the phase angle of the comets from the article by Kolokolova et al. (2004). The second value $p(\lambda)$ = 0.04 is often used for the similar study of very distant comets (Meech et al., 2009). The third value $p(\lambda)$ = 0.25 is the mean geometric albedo, which was taked from the article by Gehrz and Ney (1992), who showed that the albedo of some comets lies in the range of 0.1–0.3 on average.

To estimate the upper limit of the effective radius of the cometary nucleus, we used relation $C_d = \pi \cdot R_N^2$. The results are presented in Table 4 for the three values of the geometric albedo.

**Table 4.** The apparent magnitude of coma and upper limit of the radius of the comet nucleus. Calculations were made for three variants of mean geometric albedo: *0.04, 0.1,* and *0.25*.

| Comet | Filter | $m_d^a$ | $C_d$, [m²] | | | $R_n^*$, [km] | | |
|---|---|---|---|---|---|---|---|---|
| | | | 0.04 | 0.1 | 0.25 | 0.04 | 0.1 | 0.25 |
| C/2007 D1 (LINEAR) | R | 20.4 | 1.3·10⁹ | 5.1·10⁸ | 2.0·10⁸ | 20±6 | 13±4 | 8±2 |
| C/2007 D3 (LINEAR) | V | 22.5 | 5.2·10⁶ | 2.1·10⁶ | 8.3·10⁶ | 4±2 | 3±1 | 2±1 |
| C/2010 G3 (WISE) | R | 21.4 | 1.2·10⁸ | 4.6·10⁷ | 1.9·10⁷ | 6±3 | 4±2 | 3±2 |
| C/2010 S1 (LINEAR) | R | 18.9 | 2.5·10⁹ | 1.0·10⁹ | 4.1·10⁸ | 28±7 | 18±5 | 11±3 |
| C/2012 K6 (MgNaught) | R | 18.9 | 3.2·10⁸ | 1.3·10⁸ | 5.1·10⁷ | <10±4 | <6±3 | <4±2 |
| | V | 19.9 | 1.1·10⁸ | 4.5·10⁷ | 1.8·10⁷ | <6±2 | <4±1 | <3±1 |

a- coma magnitude in the projected annulus between $\rho_1$ =1.78 arcsec and $\rho_2$ =2.14 arcsec
*- Limits are given where cometary coma was detected.

The parameter $Af\rho$ (as a proxy for the dust production in the comet) was computed using the formula (A'Hearn et al., 1984),

$$Af\rho = \frac{4r^2\Delta^2}{\rho} 10^{0.4(m_{SUN}-m_C)},\tag{5}$$

$A$ is the albedo; $f$ is a filling factor, i.e., the ratio of the total cross section of grains; and $\rho$ is the linear radius of the field of view at the comet distance. The value $Af\rho$ is typically expressed in centimetres and is independent of the photometric aperture size. Our results are presented in Fig.3.

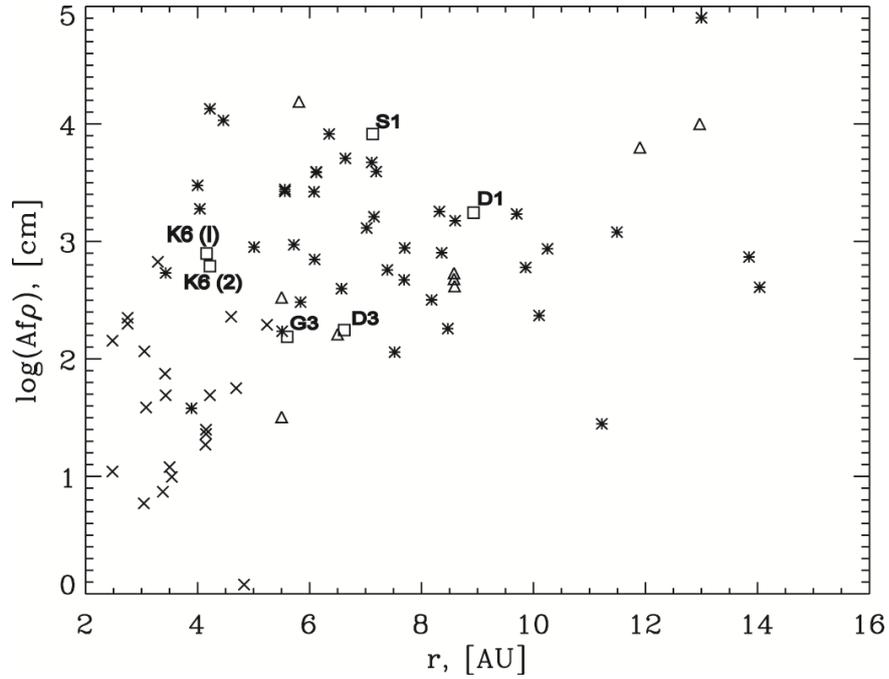

**Fig.3.** Log(*Afρ*) values obtained for target comets (squares) as the function of heliocentric distance *r* compared to the values derived for long period comets (asterisks), active centaurs (triangles) and JFC (crosses), which were active at large heliocentric distances. To compare amoung the values, we used the results presented in the papers by Lowry et al. (1999), Mazzotta Epifani et al. (2006, 2007, 2008, 2009, 2010, 2011, 2014), Meech et al. (1990, 2009), Szabó et al. (2001; 2002; 2008), Korsun et al. (2010), Rousselot et al. (2014), Shubina et al. (2014), Solontoi et al. (2012), Ivanova et al. (2014), Bauer et al. (2003) and Lara et al. (2003).

We used obtained $Af\rho$ parameters for the comets to calculate the dust mass production rate (Newburn and Spinrad, 1985; Weiler et l., 2003; Fink and Rubin, 2012). The relation to estimate the dust productivity is given by

$$Q_M = Q_N (4\pi/3) \cdot \left[ \int_{a_{min}}^{a_{max}} \rho_d(a) \cdot a^3 \cdot f(a) da \right],\tag{6}$$

where $\rho_d(a) = \rho_0 - \rho_1(a/(a+a_2))$ is the grain density (Newburn and Spinrad, 1985), with $\rho_0$=3000 kg·m$^{-3}$, $\rho_1$=2200 kg·m$^{-3}$ and $a_2$=2 μm; $f(a)$ is the differential particle size distribution, where $a$ is grain radius; $a_{min}$ and $a_{max}$ are minimal and maximal grain radii.

We used the lower and upper limits of dust grain radii equal to 5 and 1000 μm, respectively, based on the results obtained from the numerical modelling of dust environments of distant comets (Fulle 1994; Korsun 2005, 2010; Mazzotta Epifani et al. 2009). For our estimations we used two types of differential particle size distribution: in the simple form of $f(a) \sim a^{-4}$ (Korsun et al., 2005; 2010) and more complex expression of $f(a) \sim \left(1 - \frac{a_0}{a}\right)^M \cdot \left(\frac{a_0}{a}\right)^N$ (Hanner, 1983).

For our calculation we fixed the parameters, whereby the minimum grain radius $a_0$=5 μm, $M$=27 and $N$=4, which provide a peak of the function for a particle size of about 40 μm. The dust number rate is given by

$$Q_N = Af\rho \cdot [2\pi p(\lambda)\Phi(\alpha)]^{-1} \cdot \left[\int_{a_{\min}}^{a_{\max}} (f(a) \cdot a^2 / v(a))da\right]^{-1}, \quad (7)$$

where $v(a)$ is the ejection velocity of dust particle. We calculated the outflow dust velocity based on the equation found by Sekanina et al. (1992). We used the parameters based on the results obtained from numerical modeling of dust environment of some distant comets (Korsun et al., 2010, 2014; Rousselot et al., 2014), which were observed at the similar heliocentric distances as our comets. The calculated outflow velocities between 1 and 28 ms$^{-1}$ are in agreement with the numerical modeling of dust environment of some other distant active targets at the same heliocentric distances (Fulle, 1994; Mazzotta Epifani et al., 2009; Korusn et al., 2010, 2014; Rousselot et al., 2014). The numerical values of dust mass production are listed in Table 5.

**Table 5.** Dust mass production rate of the comets.

| Comet | r, [AU] | Δ, m·s$^{-1}$ | $Q_M$, kg·s$^{-1}$ | | | |
|---|---|---|---|---|---|---|
| | | | p=0.25 | | p=0.1 | |
| | | | $f(a) \sim a^{-4}$ | $f(a) \sim \left(1-\frac{a_0}{a}\right)^M \cdot \left(\frac{a_0}{a}\right)^N$ $M=27, N=4, a_0=5\mu\mu$ | $f(a) \sim a^{-4}$ | $f(a) \sim \left(1-\frac{a_0}{a}\right)^M \cdot \left(\frac{a_0}{a}\right)^N$ $M=27, N=4, a_0=5\mu\mu$ |
| C/2007 D1 (LINEAR) | 8.93 | 1-19 | 4 | 15 | 10 | 37 |
| C/2007 D3 (LINEAR) | 6.62 | 3-22 | 1 | 2 | 1 | 4 |
| C/2010 G3 (WISE) | 5.60 | 3-24 | 1 | 2 | 1 | 4 |
| C/2010 S1 (LINEAR) | 7.00 | 2-22 | 21 | 77 | 52 | 193 |
| C/2012 K6 (MgNaught) | 4.16 | 5-29 | 3 | 10 | 7 | 24 |
| | 4.22 | 5-28 | 2 | 6 | 5 | 19 |

## 4. A discussion of the dust-production mechanisms

Our observations of comets at large heliocentric distances revealed the occurrence of well-observable dust clouds near the cometary nuclei, but we did not detect any gas. Below, we discuss some mechanisms, which could lead to this.

Common cometary activity at a relatively short heliocentric distance occurs mostly due to a heating of the cometary surface by the solar radiation and subsequent sublimation of near surface water. At the large heliocentric distances, where the comets were observed, the heating is not usually sufficient to support a continuous activity. However, it can likely cause temporary eruptions of species like solid carbon monoxide or carbon dioxide, which are volatile at relatively large heliocentric distances (Meech and Svoren, 2004). The outgassing can eject a significant amount of dust from the surface. Further motion of dust grains can subsequently be influenced by solar radiation.

Alternatively, temporary outgassing and subsequent lifting of dust grains could be caused by a sudden release of internal energy, e.g. due to a phase change of amorphous ice, which can be accompanied by a release of trapped gases (Jewitt, 2009).

Despite the fact that no gas was seen, we cannot exclude that the dust particles were ejected to coma and tail by outgassing. Some molecules or ions could escape our detection. For example, $CO^+$, a potentially short-lived dissociation product, is weakly detectable in wavelengths shorter than 430 nm. Or, outgassing could happen before time of observation. Namely, we performed the spectroscopic observations of each of four comets only at a single time. The molecules and atoms that move faster than dust grains could be meanwhile gone and we only detected the dust grains, which have had not enough time to disperse. However, an intriguing aspect of this explanation is the fact that we observed, four times for four objects, only this stage of dust cloud without a gas. The most likely explanation is that at these heliocentric distances any of the emissions from the gases responsible for the activity are simply below detection limits, while the reflected solar continuum from the slowly moving dust particles in the coma is detectable.

Another mechanism to eject the dust grains from the cometary surface are the impacts of meteoroids, which blast the ejecta out of the cometary nucleus. The number density of meteoroid particles in the interplanetary space is not generally very high. However, every comet can move through the corridors of several meteoroid streams, in which the number density is considerably higher. It could be estimated that tens to hundreds of 1-gram meteoroid particles of a typical major stream can impact a comet nucleus during its passage through the stream corridor. The impact velocity ranges from several to several tens of kilometers per second, therefore the impact energy is quite high even if it is delivered by the relatively small meteoroids.

We know only the streams crossing the orbit of our planet. Almost all major streams have the parent body orbiting the Sun inside the stream corridor and the prevailing majority of these parent bodies are periodic comets. Hence, we can expect a meteoroid stream around the orbit of every periodic comet. If a given studied comet closely approaches the orbit of periodic comet, we can assume that it also passes through the corridor of associated stream. If all periodic comets were known, this assumption would enable us to estimate the number of passages of a given studied comet through the stream corridors within an appropriate time interval before the observation.

For the short-period comets 9P/Tempel 1 and 81P/Wild 2, there were found 110 and 129 approaches to the orbits of known periodic comets (Marsden and Williams, 2005) within 0.15 AU. Comet 1P/Halley approached such orbits 103 times (Ivanova et al., 2015). Considering the same sample of known, cataloged comets, we however found only 2, 13, 4, 4, and 23 approaches within 0.15 AU to the orbits of C/2007 D1, C/2007 D3, C/2010 G3, C/2010 S1, and C/2012 K6, respectively. These numbers are obviously incomplete with respect to the lack of comet discoveries with the perihelia beyond the orbit of Earth (see, e.g., Neslušan, 2007, Fig. 4). Consequently, a direct evidence of the crossings of meteoroid streams and meteoroid impacts is not possible to demonstrate and no reliable proof of the scenario that the dust coma is created by the meteoroid impacts can be done.

Sometimes an evidence supporting the particular mechanism of dust production can be gained from an image analysis showing the structure of coma and tail. Unfortunately, this kind of analysis is problematic using our images of comets in large distances. As seen in Table 1, all observations of objects at large heliocentric and, thus, geocentric distance are, necessarily, performed at a small phase angle (from 1.0º to 11.3º). Hence, we see a column of coma and tail spanning over a large interval of distances from the nucleus and the structural features, if present, are mutually over-covered. In an attempt to analyse the images, only small variations of intensity were found. Since the magnitude of these variations was smaller than the determination uncertainty, no information could be gained.

In conclusion, we cannot clearly discriminate between the mechanisms of activation of the comets in large heliocentric distances, which we observed. To achieve progress, in the future, we will need a sequence of precise observations covering a longer period to see the evolution of the dust coma of each comet. With such observations, we could see if gas is observed at a certain stage of the formation or evolution of the coma. If observations of comets are performed when they are crossing the corridors of the well-known major streams, in which the number density of meteoroids can be rather reliably estimated, this would help to support or reject the meteoroid-impact hypothesis.

5. Results

(1) No gas emissions were detected in the spectra of four distant comets: C/2007 D1 (LINEAR), C/2007 D3 (LINEAR), C/2010 G3 (WISE), and C/2010 S1 (LINEAR). The upper limits of the gas production rates for the main neutral molecules in the cometary comae were calculated.

(2) The results of the photometric investigation of five distant comets presented here show activity properties that are typical for new comets. The Afρ value obtained (Fig.3) and the dust production (Tab. 5) for our objects are of the same order as those measured for most of dynamically new and long period comets (Mazzotta Epifani et al. 2009; 2010; Meech et al. 2009; Szabó et al., 2001; 2002; 2008; Korsun et al., 2010; Ivanova et al., 2014; Rousselot et al., 2008). Only comet C/2010 S1 (LINEAR) features a higher level of the dust production rate than the other objects. Shubina et al. (2014) estimated the comet dust production to be between 20 and 60 kg s$^{-1}$ at the distance of 6.3 AU. The activity of the comet is similar to that of objects 29P/Schwassmann-Wachmann 1 (Szabó et al., 2002) and 174P/Echeclus (Rousselot et al., 2013).

(3) Estimates of the upper limits of effective cometary radii yield the values (Tab. 4) that are similar to those of a majority of comets (Lamy et al., 2004). Our results depend on particular values of the albedo used for the calculations. We found the radii from 6 to 28 km, 4 to 13 km, and 3 to 8 km using albedo 0.04, 0.1, and 0.25, respectively.

**Acknowledgements**

This work was supported in part by the Slovak Grant Agency VEGA (grants No. 2/0031/14 and 2/0032/14) and the implementation of the project SAIA. Grateful acknowledgement is made to Dr. Lukyanyk for discussion and constructive criticism about programming methods under the IDL. The results are based on the observations made with the 6-m telescope of SAO RAN and the 60-cm Zeiss 600 telescope of the Peak Terskol Observatory (IC AMER).